\documentclass[aps,prd,groupedaddress,graphicx,nofootinbib]{revtex4}
\usepackage{amsmath,amssymb,graphics,graphicx,color,epsf}
\usepackage{subfigure}
\usepackage[scanall]{psfrag}  
\usepackage{tikz}

\begin{document}

\title{Testing hilltop supernatural inflation with gravitational waves}

\author{Chia-Min Lin}

\affiliation{Fundamental Education Center, National Chin-Yi University of Technology, Taichung 41170, Taiwan}

\date{Draft \today}

\begin{abstract}
The scale of small-field inflation cannot be constrained via primordial gravitational waves through measurement of tensor-to-scalar ratio $r$.
In this study, I show that if cosmic strings are produced after symmetry breaking at the end of hilltop supernatural inflation, this small-field inflation model can be tested through the production of gravitational waves from cosmic strings. Future experiments of gravitational wave detectors will determine or further constrain the parameter space in the model.
\end{abstract}
\maketitle
\large
\baselineskip 18pt
\section{Introduction}
The observation of gravitational waves \cite{TheLIGOScientific:2016wyq, Abbott:2017xzu} not only confirms again the validity of Einstein's general theory of relativity but also heralds a new era of cosmology and astrophysics. More precise measurements would be provided by future experiments such as LIGO \cite{TheLIGOScientific:2014jea}, Virgo \cite{TheVirgo:2014hva}, LISA \cite{Audley:2017drz}, DICIGO/BBO \cite{Yagi:2011wg}, the Einstein Telescope (ET) \cite{Punturo:2010zz}, Cosmic Explorer (CE) \cite{Evans:2016mbw}, Taiji \cite{Guo:2018npi}, and TianQin \cite{Mei:2020lrl}. 

It is interesting and important to investigate the ramification of the observation of gravitational waves to cosmology, in particular, inflationary cosmology \cite{Starobinsky:1980te, Sato:1980yn, Guth:1980zm, Linde:1981mu}. Inflation models can be roughly divided into large-field and small-field models. There are two major observables that can be used as the first test for inflation models, namely the spectral index $n_s$ and the tensor-to-scalar ratio $r$. In the case of large-field models, we have the inflaton field $\psi > M_P$ during inflation, where $M_P=2.4 \times 10^{18}\mbox{ GeV}$ is the reduced Planck mass. A representative model is chaotic inflation \cite{Linde:1983gd} with a quadratic potential of the inflaton field $\psi$
\begin{equation}
V=\frac{1}{2}m^2_\psi \psi^2.
\label{eq1}
\end{equation}
A large-field inflation model typically predicts a high energy scale during inflation and may produce observable primordial gravitational waves which can be constrained experimentally through the tensor-to-scalar ratio $r$.  The model of Eq.~(\ref{eq1}) is simple and elegant. However, it predicts a large $r$ and is being ruled out by experimental results (see Fig. 8 of \cite{Akrami:2018odb}). This also shows the observation of primordial gravitational waves is a good method to test large-field inflation models. On the other hand, there are small-field inflation models that corresponds to $\psi < M_P$. These models have a smaller energy scale during inflation. Theoretically, they also produce primordial gravitational waves, but the value of $r$ is typically too small to be observed in near-future experiments. A renowned example of small-field inflation model is hybrid inflation \cite{Linde:1993cn} which will be described in the next section. One character of this model is that there could be a symmetry breaking after inflation with possible topological defects produced depend on the symmetry which is broken. Examples are monopoles, domain walls, or cosmic strings. Production of monopoles or domain walls usually is fatal to a model building. Indeed one motivation to introduce the idea of cosmic inflation is to dilute unwanted relics produced in the early universe. On the other hand, cosmic strings production provides a palatable possibility. 

Cosmic strings emit gravitational waves.
Gravitational waves from cosmic strings has been used to test non-standard cosmologies \cite{Cui:2017ufi, Gouttenoire:2019kij, Gouttenoire:2019rtn}, probe the pre-BBN universe \cite{Cui:2018rwi}, test grand unified theories \cite{King:2020hyd, Buchmuller:2019gfy, Chakrabortty:2020otp, Chigusa:2020rks}, probe axion physics \cite{Ramberg:2019dgi}, test seesaw and leptogenesis \cite{Dror:2019syi, Samanta:2020cdk},  test cosmic string formation during inflation \cite{Guedes:2018afo, Cui:2019kkd}, probe primordial black holes \cite{Jenkins:2020ctp}, probe quintessential inflation \cite{Bettoni:2018pbl}, and test baryon and lepton number violation \cite{Fornal:2020esl}.
In this paper, I propose the idea to probe the scale of (hilltop) hybrid inflation, via gravitational waves from the cosmic strings which are produced after inflation.

\section{Hilltop Supernatural Inflation}
\label{sec2}
The potential for a hybrid inflation \cite{Linde:1993cn} is given by
\begin{equation}
V=\frac{1}{2}m^2_\psi \psi^2 + g^2\psi^2\phi^2+\kappa^2(\phi^2-\Lambda^2)^2,
\label{eq2}
\end{equation}
where $\psi$ is the inflaton field and $\phi$ is the waterfall field. During inflation, the field value of $\psi$ gives a large positive mass to $\phi$ therefore it is trapped to $\phi=0$ and the potential during inflation is of the form
\begin{equation}
V=V_0+\frac{1}{2}m^2_\psi \psi^2,
\label{eq3}
\end{equation}
where $V_0=\kappa^2 \Lambda^4$. The end of inflation is determined by $m^2_\phi=0$ when the waterfall field starts to become tachyonic which implies
\begin{equation}
\psi_{end}=\frac{\sqrt{2}\Lambda \kappa}{g}.
\label{eq4}
\end{equation}

The effective constant $V_0$ during inflation in Eq.~(\ref{eq3}) is crucial to make the model a small-field inflation. If $V_0 \ll (1/2)m^2_\psi \psi^2$, it becomes a large-field model of Eq.~(\ref{eq1}). Therefore it is assumed that $V_0 \gg (1/2)m^2_\psi \psi^2$. However, there can be variations of $V_0$ up to several orders which cannot be determined by current experimental results. In particular, as mentioned in the previous section, $V_0$ is not expected to be constrained through tensor to scalar ratio $r$. Theoretically, one possibility to decide $V_0$ is to motivate it through high energy physics theories. In supernatural inflation \cite{Randall:1995dj}, $V_0$ is proposed to be determined from a supersymmetry breaking scale, namely $V_0=M_S^4$ where $M_S=\sqrt{m_SM_P}$ is the gravity-mediated SUSY breaking scale and $m_S\sim m_\psi$ is the soft mass. Depends on $m_\psi \sim \mathcal{O} (1-100) \mbox{ TeV}$, $V_0$ lies in the range $V_0 \sim (10^{11-12} \mbox{ GeV})^4$. If the gravitino mass (which is roughly the same order as soft mass \cite{Dine:1995kz, Enqvist:2003gh}) is about $100$ TeV, there can be a nonthermal origin for dark matter production \cite{Moroi:1999zb, Acharya:2008bk, Acharya:2009zt, Acharya:2010af, Kane:2011ih, Kim:2016spf}. The potential energy scale $V_0 \sim (10^{12} \mbox{ GeV})^4$ has been used for baryogenesis \cite{Lin:2020lmr} and constructions of SUSY unified models in hilltop supernatural inflation \cite{Kohri:2013gva}. Alternatively, if the model of hybrid inflation is constructed via grand unified theories (GUT), $V_0$ is a parameter in the GUT potential, which is not necessarily related to the SUSY breaking.  

Hybrid inflation and its supernatural version are elegant and simple models, unfortunately they have been ruled out by experimental observation \cite{Akrami:2018odb}. The reason is that due to the concave upward potential in Eq.~(\ref{eq3}), it predicts a blue spectral index $n_s>1$ for primordial density perturbation. One method to obtain a red spectral index $n_s<1$ is to introduce an additional term in the potential and this makes the model to become a hilltop inflation \cite{Lin:2009yt}. One of the possibilities is to consider the following potential
\begin{equation}
V(\psi)=V_0+\frac{1}{2}m_\psi^2 \psi^2 - \lambda \psi^4 \equiv V_0 \left( 1+\frac{1}{2}\eta_0 \frac{\psi^2}{M_P^2}\right)-\lambda \psi^4,
\end{equation}
where $\eta_0 \equiv m^2_\psi M_P^2/V_0$. This model is reasonably easy that analytical results for observables such as the spectral index $n_s$ and tensor-to-scalar ratio $r$ can be obtained under slow-roll approximation \cite{Kohri:2007gq}. One attractive feature of this model is that the additional quartic term can be obtained through a SUSY breaking A-term with the right magnitude of $n_s$ in order to have $n_s=0.96$ which fits cosmological observations. By fixing $n_s=0.96$ and imposing CMB normalization $P^{1/2}_R=5 \times 10^{-5}$, the inflaton field value when our observable universe exit the horizon (at the number of e-folds chosen to be $N=60$) is given by
\begin{equation}
\left( \frac{\psi}{M_P} \right)^2 = \left( \frac{V_0}{M_P^4} \right) \frac{\eta_0+0.02}{12 \lambda},
\end{equation}
 and the inflaton field value at the end of inflation is
\begin{equation}
\psi^2_{end}=\frac{3\eta_0}{(2\eta_0-0.02)e^{120\eta_0}+\eta_0+0.02}\psi^2.
\label{eq7}
\end{equation}
The above equation would be useful in subsequent sections.

\section{Cosmic Strings}

If the waterfall field $\phi$ in Eq.~(\ref{eq2}) is a real field, it possesses a $Z_2$ symmetry which is broken after inflation when the field obtains a vacuum expectation value $\phi=\Lambda$ and there will be formation of domain walls as topological defects. In (hilltop) supernatural inflation, $\phi$ is a complex superfield and the last term in Eq.~(\ref{eq2}) is modified to 
\begin{equation}
\kappa^2\left( |\phi|^2-\Lambda^2 \right).
\end{equation}
If there is a $U(1)$ invariance realized by introducing a vector field $A_\mu$ with the transformations
\begin{equation}
\phi \rightarrow \phi e^{i f(x)},    \;\;\;\;\;   A_\mu \rightarrow A_\mu - \frac{1}{e}\partial_\mu f(x),
\end{equation}
where $f(x)$ is a real function, there will be production of (local) cosmic strings after symmetry breaking (see \cite{Hindmarsh:1994re} for review of cosmic strings). It has been shown that production of cosmic string in SUSY GUTs is a quite generic result \cite{Jeannerot:2003qv}. The energy per unit length of the string (also known as the string tension) is \cite{Kibble:1976sj}
\begin{equation}
\mu \simeq 2 \pi \Lambda^2.
\label{eq10}
\end{equation}
This value is exact in the case $2\kappa/e^2=1$ \cite{Bogomolny:1975de}.
String tension is commonly expressed via a dimensionless combination $G\mu$ in literatures, where $G$ is Newton's constant. Cosmic strings can also contribute to primordial density perturbation with a different shape of the spectrum from that of inflaton field fluctuations. Current constraint from CMB measurement is $G\mu < 1.1 \times 10^{-7}$ \cite{Charnock:2016nzm}. As we can see in the next section, gravitational waves measurement gives a stronger constraint to $G\mu$.

\section{Gravitational waves from cosmic strings}

Cosmic strings have tension and the corresponding energy densities, they evolve and produce gravitational radiation when they oscillate \cite{Hindmarsh:1994re}. The spectrum of the background gravitational radiation produced by cosmic strings is obtained in \cite{Vilenkin:1981bx,Hogan:1984is,Vachaspati:1984gt, Accetta:1988bg, Bennett:1990ry, Blanco-Pillado:2017oxo}. 

Cosmic strings form through Kibble mechanism \cite{Kibble:1976sj} during a phase transition associated with spontaneous symmetry breaking. There is roughly one long string in each Hubble patch when the formation occurs. After the production of cosmic strings, long strings intercommute to from closed string loops. When the damping with the thermal plasma comes to an end, cosmic strings enter the phase of scaling evolution and oscillate. The initial length of a string loop created at time $t_i$ is 
\begin{equation}
l_i=\alpha t_i,
\end{equation} 
where $\alpha \simeq 0.1$ is a loop size parameter \cite{Blanco-Pillado:2013qja}. After formation, string loops emit gravitational radiation at a constant rate
\begin{equation}
\frac{dE}{dt}=-\Gamma G \mu^2,
\end{equation}
where $\Gamma \simeq 50$ \cite{Blanco-Pillado:2017rnf} from Nambu-Goto simulations. Thus $l_i$ decreases as 
\begin{equation}
l(t)=\alpha t_i-\Gamma G \mu (t-t_i).
\end{equation}

I assume here that local cosmic strings mainly decay via production of stochastic background of gravitational waves as suggested in \cite{Matsunami:2019fss, Moore:1998gp, Olum:1999sg, Moore:2001px}\footnote{This issue is controversial, opposite result was claimed in \cite{Hindmarsh:2017qff, Vincent:1997cx}. For global strings, the main decay channel is the production of massless Goldstone boson \cite{Vilenkin:2000jqa}.}.
The gravitational wave density per unit frequency observed today is (See appendix B of \cite{Gouttenoire:2019kij} for a derivation of this formula.)
\begin{equation}
\Omega_{GW}=\frac{f}{\rho_c}\frac{d\rho_{GW}}{df}=\sum_k \Omega^{(k)}_{GW}(f),
\end{equation}
with
\begin{equation}
\Omega^{(k)}_{GW}(f)=\frac{1}{\rho_c}\frac{2k}{f}\frac{(0.1)\Gamma_k G\mu^2}{\alpha(\alpha+\Gamma G \mu)}\times \int_{t_f}^{t_0}d\tilde{t}\frac{C_{eff}(t_i)}{t^4_i}\left[ \frac{a(\tilde{t})}{a(t_0)} \right]^5 \left[ \frac{a(t_i)}{a(\tilde{t})} \right]^3 \Theta(t_i-t_F), 
\end{equation}
where $\rho_{GW}$ is the energy density of gravitational waves, $f$ is the frequency today, and $\rho_c=3H^2_0/8\pi G$ is the critical density, $\Gamma_k=\Gamma/(3.60 k^{4/3})$, $t_F$ is the time when the scaling regime of the string network has been achieved, $\Theta$ is the Heaviside theta function, and $C_{eff}=0.39$ (5.4) for matter domination (radiation domination) is the loop-formation efficiency obtained by solving the Velocity-dependent One-Scale (VOS) equations. Current limit from the European Pulsar Timing Array (EPTA) implies $G\mu \lesssim 10^{-11}$ \cite{vanHaasteren:2011ni}. Future experiments have the potential to reduce the limit up to $G \mu \lesssim 10^{-19}$ or so (for example, see Figure 4 in \cite{Gouttenoire:2019kij}).

\section{Testing hilltop supernatural inflation}

\begin{figure}[t]
  \centering
\includegraphics[width=0.6\textwidth]{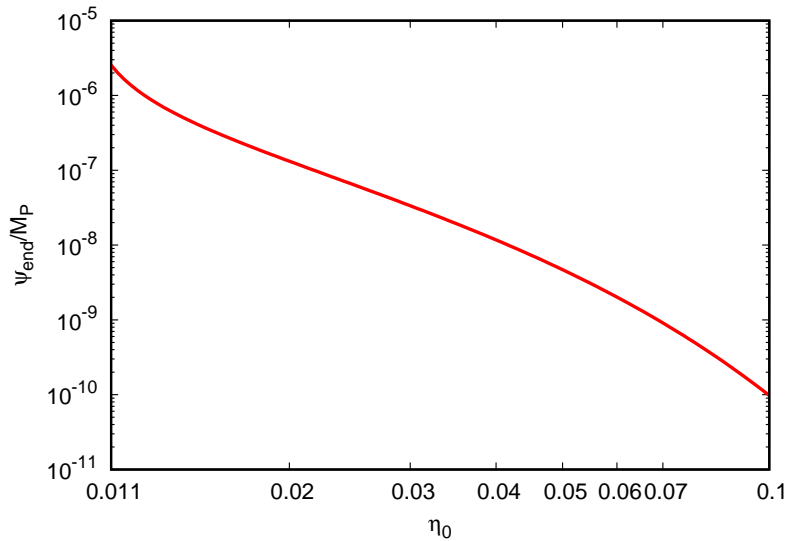}
  \caption{$\psi_{end}/M_P$ as a function of $\eta_0$ for $V_0=10^{-24}M_P^4$.}
  \label{fig1}
\end{figure}

\begin{figure}[t]
  \centering
\includegraphics[width=0.6\textwidth]{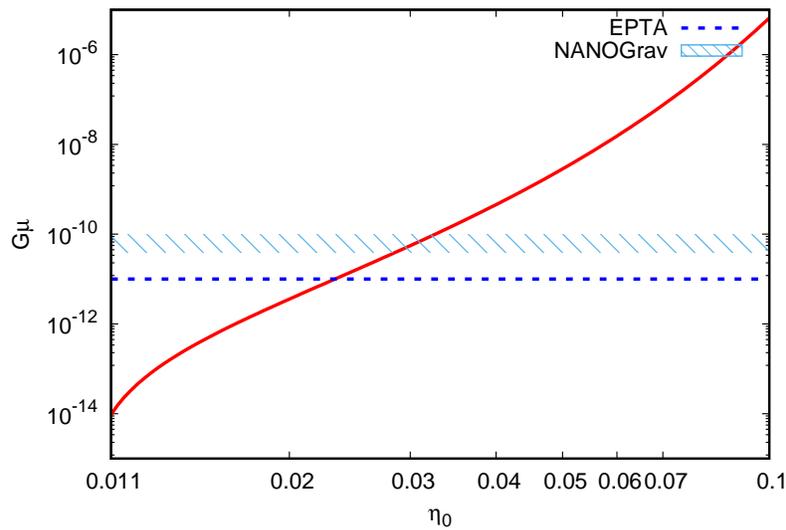}
  \caption{$G\mu$ as a function of $\eta_0$ for $V_0=10^{-24}M_P^4$ and $g=1$. Current upper limit from the European Pulsar Timing Array (EPTA) and possible signal from NANOGrav are also shown in the figure.}
  \label{fig2}
\end{figure}

There is a lower bound for $\Lambda$ in our parametrization which can be obtained by imposing $g<1$ to avoid the model becoming non-perturbative. By using Eq.~(\ref{eq4}) and $V_0=\kappa^2 \Lambda^4$, we have
\begin{equation}
\Lambda=\frac{\sqrt{2V_0}}{g\psi_{end}},
\label{eq16}
\end{equation}
where $\psi_{end}$ is given by Eq.~(\ref{eq7}). Supposing $V_0=\kappa^2 \Lambda^4 \sim (10^{12} \mbox{ GeV})^4 \sim 10^{-24}M_P^4$ as motivated in Section \ref{sec2}, I plotted $\psi_{end}$ in Fig.~\ref{fig1}. From Eq.~(\ref{eq16}), the condition $g<1$ gives a lower bound for $\Lambda$. As can be seen in Eq.~(\ref{eq10}), the lower bound of $\Lambda$ gives a lower bound of $G\mu$. I plotted $G\mu$ as a function of $\eta_0$ in Fig.~\ref{fig2} for $g=1$. For $g<1$ the curve of $G\mu$ in the plot would shift upward. On the other hand, if $V_0<10^{-24}M_P^4$, the corresponding curve would shift downward. As can be seen from the figure, the current limit from EPTA gives a constraint of $\eta_0 \lesssim 0.02$. This data would be improved by several orders in the near future as described in the introduction section. Recently there is a possible signal of stochastic gravitational waves background from NANOGrav Collaboration \cite{Arzoumanian:2020vkk}. This corresponds to a string tension $G\mu \in (4\times 10^{-11}, 10^{-10})$ at the $68 \%$ confidence level \cite{Ellis:2020ena}\footnote{In \cite{Blasi:2020mfx}, a slightly different value of $G\mu \in (6\times 10^{-11}, 1.7\times 10^{-10})$ at the $68 \%$ confidence level is obtained.}. If true, this already gives $\eta_0 \sim 0.03$ in our model as can be seen from Fig.~\ref{fig2}. Future experiments would be able to further clarify these results.

If eventually cosmic strings are confirmed to be discovered, parameters like $V_0$ and $\eta_0$ can be constrained. On the other hand, if there is no discovery of cosmic strings in the future, models that predict the existence of cosmic strings at certain energy scales would be ruled out. Note that $V_0$ is small in small field inflation, which is not expected to be determined through primordial gravitational waves (namely tensor-to-scalar ratio $r$). It is interesting that it is possible to probe $V_0$ through gravitational waves from cosmic strings in this kind of model. 

It is possible to build a hilltop supernatural inflation model in the framework of SUSY unified theories \cite{Kohri:2013gva}. One interesting possibility is to investigate cosmic strings produced from the symmetry breaking of $SU(5)\times U(1)$ (flipped $SU(5)$) to the Standard Model gauge group with the matter parity $G_{SM} \times Z_2$. In addition, from the point of view of connecting this model to supersymmetric particle physics, $\eta_0$ is determined  from a soft mass in a flat-direction. Therefore measurement of $\eta_0$ allows us to probe high energy physics theory through observations of gravitational waves.

\section{Conclusion and Discussion}
\label{con}

Traditionally inflation models are tested via an $n_s$-$r$ plot of the spectral index and tensor to scalar ratio. These data more or less stimulate the study of large field inflation models because of the prospect of detecting primordial gravitational waves. On the other hand, small field inflation models produce negligible primordial gravitational waves and cannot be constrained through $r=16 \epsilon$, where $\epsilon \equiv (1/2)M_P^2(V^\prime/V)^2$ is a slow-roll parameter. This can be seen from the equation of number of e-folds:
\begin{equation}
N=\frac{1}{M_P}\int_{\psi_{end}}^{\psi}\frac{d\psi}{\sqrt{\epsilon}} \simeq \frac{\frac{\Delta \psi}{M_P}}{\sqrt{\epsilon}}.
\end{equation} 
For small field inflation models $\Delta \psi \ll M_P$, therefore, in order to obtain enough number of e-folds, $\epsilon$, hence $r$ has to be small. This is essentially the argument that gives the Lyth bound \cite{Lyth:1996im}. 

In this study, I point out a method to investigate and distinguish an important class of small field inflation models, namely hilltop supernatural inflation, which may be connected to ideas from particle physics such as supersymmetry or grand unified theories. Cosmic strings are typical and characteristic by-product of many hybrid inflation models. The measurement of gravitational waves produced from cosmic strings could be regarded as a complementary way to the observation of primordial gravitational waves, which is tailor-made for this kind of small field inflation models. In the near future, as new data are collected, we can certainly learn more about these models. One may be tempted to say that \textit{there is plenty of room at the bottom} of the $n_s$-$r$ diagram! 

\acknowledgments
This work is supported by the Ministry of Science and Technology (MOST) of Taiwan under Grant No. MOST 109-2112-M-167-001.

\end{document}